\pdfoutput=1
\documentclass{egpubl}
\usepackage{eurova2023}

 \WsPaper           %

 \electronicVersion %

\ifpdf \usepackage[pdftex]{graphicx} \pdfcompresslevel=9
\else \usepackage[dvips]{graphicx} \fi

\PrintedOrElectronic

\usepackage{t1enc,dfadobe}

\usepackage{egweblnk}
\usepackage{cite}

\graphicspath{{figures/}{pictures/}{images/}{./}} %

\usepackage{array}
\usepackage{amsfonts}
\usepackage{amsmath}
\usepackage{amssymb}
\usepackage[inline]{enumitem}
\usepackage{multirow}
\usepackage{multicol}
\usepackage{wrapfig}
\usepackage{caption}
\usepackage{subcaption}
\usepackage{makecell}

\title{ShaRP: Shape-Regularized Multidimensional Projections}

\def\*#1{\mathbf{#1}}

\author[A. Machado \& A. Telea \& M. Behrisch]
{\parbox{\textwidth}{\centering A.\,Machado$^1$\orcid{0000-0002-1129-4628}
        , A. Telea$^{1}$\orcid{0000-0003-0750-0502}
        , M. Behrisch$^{1}$\orcid{0000-0002-1102-103X}
        }
        \\
{\parbox{\textwidth}{\centering $^1$Department of Information and Computing Science, Utrecht University, Netherlands
       }
}
}

\begin{document}

\maketitle

\begin{abstract}

Projections, or dimensionality reduction methods, are techniques of choice for the visual exploration of high-dimensional data. Many such techniques exist, each one of them having a distinct \emph{visual signature} --- \emph{i.e.}, a recognizable way to arrange points in the resulting scatterplot. Such signatures are implicit consequences of algorithm design, such as whether the method focuses on local \emph{vs} global data pattern preservation; optimization techniques; and hyperparameter settings. We present a novel projection technique --- ShaRP --- that provides users explicit control over the visual signature of the created scatterplot, which can cater better to interactive visualization scenarios. ShaRP scales well with dimensionality and dataset size, generically handles any quantitative dataset, and provides this extended functionality of controlling projection shapes at a small, user-controllable cost in terms of quality metrics.



\end{abstract}

\renewcommand{\baselinestretch}{0.95}

\section{Introduction}
\label{sec:Introduction}

Projection, also called Dimensionality Reduction (DR), methods are popular tools for exploring high-dimensional datasets. They transform the task of discovering data  patterns in high-dimensional spaces into a perceptually-driven search and inspection task of \emph{visual patterns} in 2D or 3D through scatterplots. Prior research has shown that such scatterplots help uncovering topological aspects, such as groupings, outliers, and correlations in the data\cite{quality-metrics-infovis,DBLP:conf/chi/PandeyKFBB16,DBLP:journals/tvcg/WangFCZFSYC18}.

However, visual patterns in a projection depend not only on the underlying \emph{data}, but also on how the DR technique is designed. For example, for the same dataset, t-SNE tends to create organic, round, structures; Auto-Encoders create starburst-like clusters; and UMAP creates very dense, round, clusters, to mention just a few\,\cite{nonato18,EspadotoDRSurvey}. We further call such aspects the \emph{visual signature} of a projection technique.

We believe that users can benefit from having \emph{direct} control over the visual signatures of a projection technique. For instance, when performing interactive data labeling using rectangular selections or displaying image thumbnails over data clusters (see \autoref{fig:squaredness-overlaid-mnist}), a projection whose clusters resemble rectangles would be more suitable than one creating various-shaped clusters (if all other aspects of the two projections, \emph{e.g.}, quality, are similar). However, controlling such visual signatures is typically hard with current projection methods.

To fill that gap, we present \emph{ShaRP} (standing for \textbf{Sha}pe \textbf{R}egularized Neural \textbf{P}rojection), to the best of our knowledge the first algorithm that provides users with direct control over cluster shapes in their projection scatterplots. We next describe the technique, illustrate this new shape regularization ability, show that it comes at a user-controlled penalty to standard quality metrics, and point towards avenues for further exploration.

\section{Background and Related Work}
\label{sec:RelatedWork}

We first introduce a few notations: A dataset $\*X = \{\*x_i\}_{i=1,\dots,m}$ has  $m$ samples $\*x_i = [x_{i1}, \ldots, x_{in}]^T$, where $\*x_i$ is a point in $\mathbb{R}^n$ with components $x_{ij}$, $1 \leq j \leq n$ and an optional label $y_i \in \{1, \dots, K\}$. We use capitals to denote the set of all elements for the corresponding small letter, \emph{e.g.}, $\bar{Y} = \{\bar{y}_i\}_{i=1,\dots,m}$. We denote the Euclidean norm by $\lVert \*x \rVert_2 = \sqrt{\*x^T \*x}$ and
the expected value of a function of a random variable $\*z$ distributed according to $p$ by $\mathbb{E}_{\*z \sim p}[f(\*z)]$. Further, we use $\theta$ to denote probability distribution parameters, for example $\theta = (\vec{\*\mu} \in \mathbb{R}^2, \vec{\*\sigma}^2 \in \mathbb{R}^2)$ for a 2D Diagonal Gaussian distribution.

\smallskip
\noindent\textbf{Dimensionality reduction:} Projection algorithms are formally functions
$
    P_\eta : \mathbb{R}^n \to \mathbb{R}^q
$
where $q \ll n$ and $\eta$ denote (hyper)parameters.
In this work we focus on 2D scatterplots ($q=2$) and use the term ``projections'' to refer to both such 2D scatterplots and the DR algorithms that create them.

Many projection algorithms are available nowadays. These are described from technical perspectives (how they differ design-wise) in several surveys\,\cite{maaten09_survey,hoffman02,yin07_survey,bunte11,sorzano14_survey,cunningham15_survey,maljovec15}; and from the perspective of \emph{local} quality metrics\,\cite{nonato18,EspadotoDRSurvey}. Well-known projection methods include Principal Component Analysis (PCA) --- a simple, easy to code, but qualitatively limited method especially for complex non-planar data structures embedded in high dimensions\,\cite{pca}; Isomap --- a technique which works well if the data resides on a (single) high-dimensional manifold\,\cite{isomap}; t-SNE, which works well for arbitrary high-dimensional data distributions but has challenges in controlling (and predicting) the shapes of the emerging visual clusters\,\cite{tsne,barneshut-tsne,Ulyanov2016}; and UMAP, similar to t-SNE in terms of ease and of visual cluster control\,\cite{DBLP:journals/corr/abs-1802-03426}.

\smallskip 
\noindent\textbf{Visual signatures:} Significant work aimed to develop ways to control or adapt the visual signatures of projections\,\cite{hagrid, hypernp, DBLP:journals/corr/MakhzaniSJG15} and studying whether it is \emph{possible} to do so for existing projection methods, as follows. 
Cutura et al.\,\cite{hagrid} use space-filling curves to adapt the position of data points in image thumbnail scatterplots such that they are non-overlapping. This idea is effective but limited to image datasets, while our proposed technique is designed to be generic.
The perplexity parameter in t-SNE is responsible for, among other things, the visual appearance of the projection. Its effect is intricately enmeshed within t-SNE, leading to cluster shapes, sizes, and distances that do not necessarily convey meaning\,\cite{tsne-perplexity}. Our technique minimizes the variability of this visual appearance.
Since fully doing away with hyperparameters might be infeasible, techniques such as HyperNP\,\cite{hypernp} learn to simulate their effects on the resulting projection. Our technique instead works well for several datasets using a single setting of hyperparameters, reducing the need for such simulation.
Makhzani \emph{et al.}\,\cite{DBLP:journals/corr/MakhzaniSJG15} propose an approach similar to ours. They adapt an auto-encoder into an adversarial setting so as to increase the quality of the projection (\emph{e.g.}, better cluster separability) that would have been created by the vanilla auto-encoder. Our method instead aims mainly at putting cluster shapes under the user's control. %

\section{ShaRP: Shape-Regularized Neural Projection}
\label{sec:Method}

We now introduce ShaRP, our novel DR technique, which is based on deep neural networks. Such networks are, in general, able to approximate complex non-linear functions and 
have several desirable features that ShaRP inherits:

\smallskip
\noindent\textbf{Scalable:} ShaRP scales linearly in the number of samples because it avoids precomputing pairwise distances or covariance matrices, like in PCA or t-SNE, and lends itself to hardware-acceleration through GPUs or TPUs tailored for fast deep learning.
    
\smallskip
\noindent\textbf{Parametric:} ShaRP operates in a ``learn once, project as needed'' fashion. It learns to parameterize a projection function instead of only outputting the projected points, such as t-SNE or UMAP. This allows ShaRP to project data it was not trained on along with existing data (out-of-sample ability).
    
\smallskip
\noindent\textbf{Generic:} ShaRP handles any dataset comprised of numeric features and can be applied to a wide range of datasets using the same or only slightly adapted hyperparameter settings.

\smallskip
\noindent\textbf{Sound:} ShaRP scores comparably to state of the art techniques in relevant projection quality metrics.

\noindent To these, ShaRP adds two flavors of \textbf{Shape Regularization}:
\begin{itemize}
    \item\emph{Intra-projection}: ShaRP creates point clusters having shapes coming from the same \emph{family}: ellipses, rectangles, triangles.
    \item\emph{Inter-projection}: Running ShaRP over different datasets produces a consistent visual signature where differences in the projections are driven mainly by data patterns. %
\end{itemize}

ShaRP is implemented in Python using Keras (Tensorflow backend)\,\cite{chollet2015keras}, Tensorflow Probability\,\cite{TensorflowProbability} for sampling and calculating log-probabilities under different distributions and is publicly available at \URL{https://github.com/amreis/sharp}.

\subsection{Method description}
\label{subsec:the-sharp-method}
ShaRP belongs to the family of Representation Learning \,\cite{BengioRepresentationLearningSurvey} techniques, \emph{i.e.}, it learns a \emph{latent encoding} for input data. A latent encoding is a vector $\*r \in \mathbb{R}^q$, where $\*r = f(\mathbf{x})$ is a low-dimensional representation of the input $\*x \in \mathbb{R}^n$ that enables a reconstruction of $\*x$ with minimal errors. As said earlier, we aim to create 2D projections, so $q = 2$.

\begin{figure}[htbp]
    \centering
    \begin{subfigure}{0.31\linewidth}
        \centering
        \includegraphics[trim=100 100 100 100, clip, keepaspectratio, width=\linewidth]{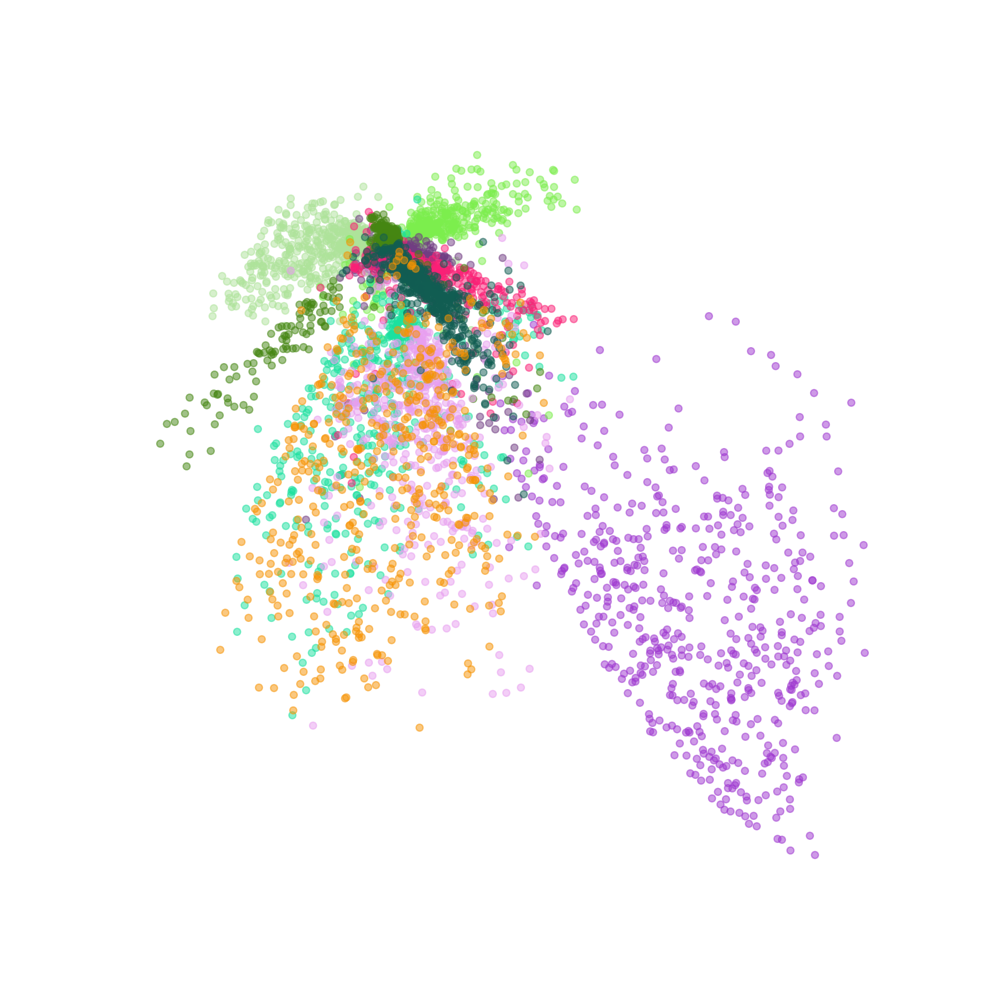}
        \caption{AE (0.54)}
        \label{fig:inset-ae-proj}
    \end{subfigure}
    \hfill
    \begin{subfigure}{0.31\linewidth}
        \centering
        \includegraphics[trim=100 100 100 100, clip, keepaspectratio, width=\linewidth]{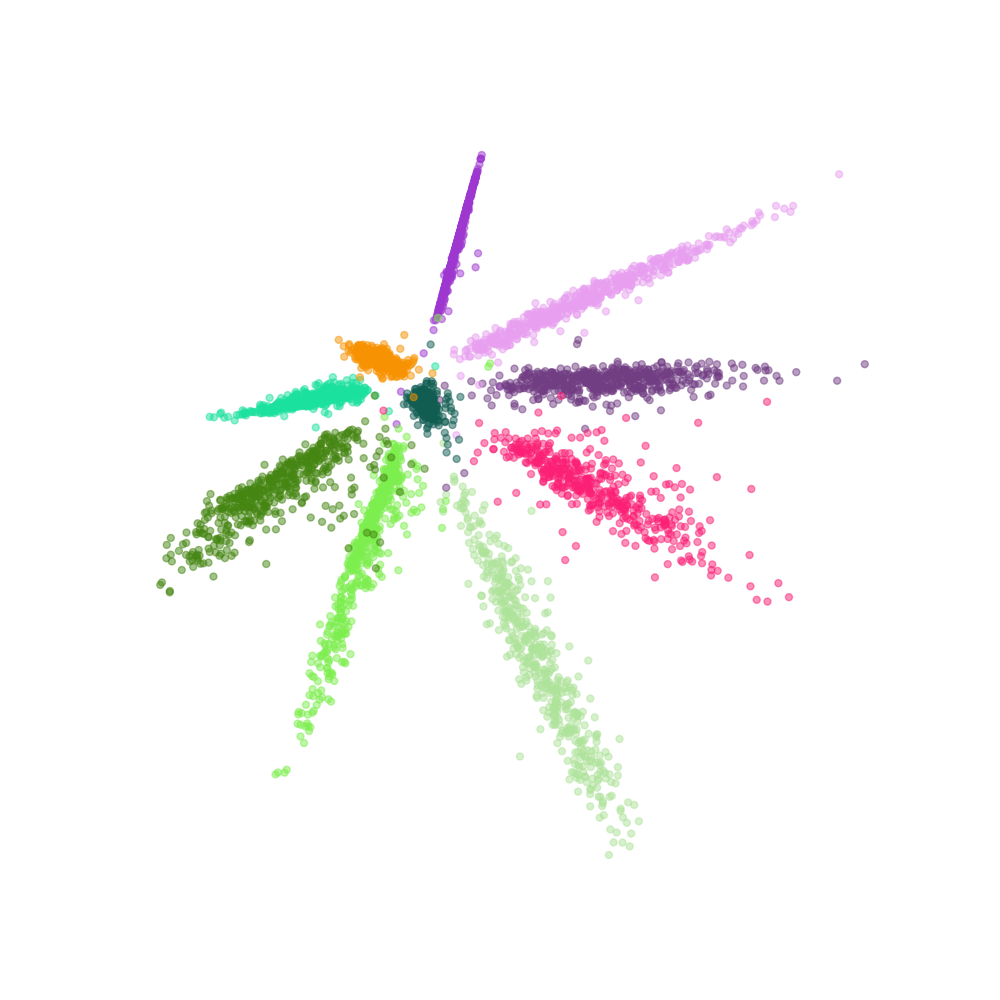}
        \caption{SSNP (0.87)}
        \label{fig:inset-ssnp-proj}
    \end{subfigure}
    \hfill
    \begin{subfigure}{0.31\linewidth}
        \centering
        \includegraphics[trim=100 100 100 100, clip, keepaspectratio, width=\linewidth]{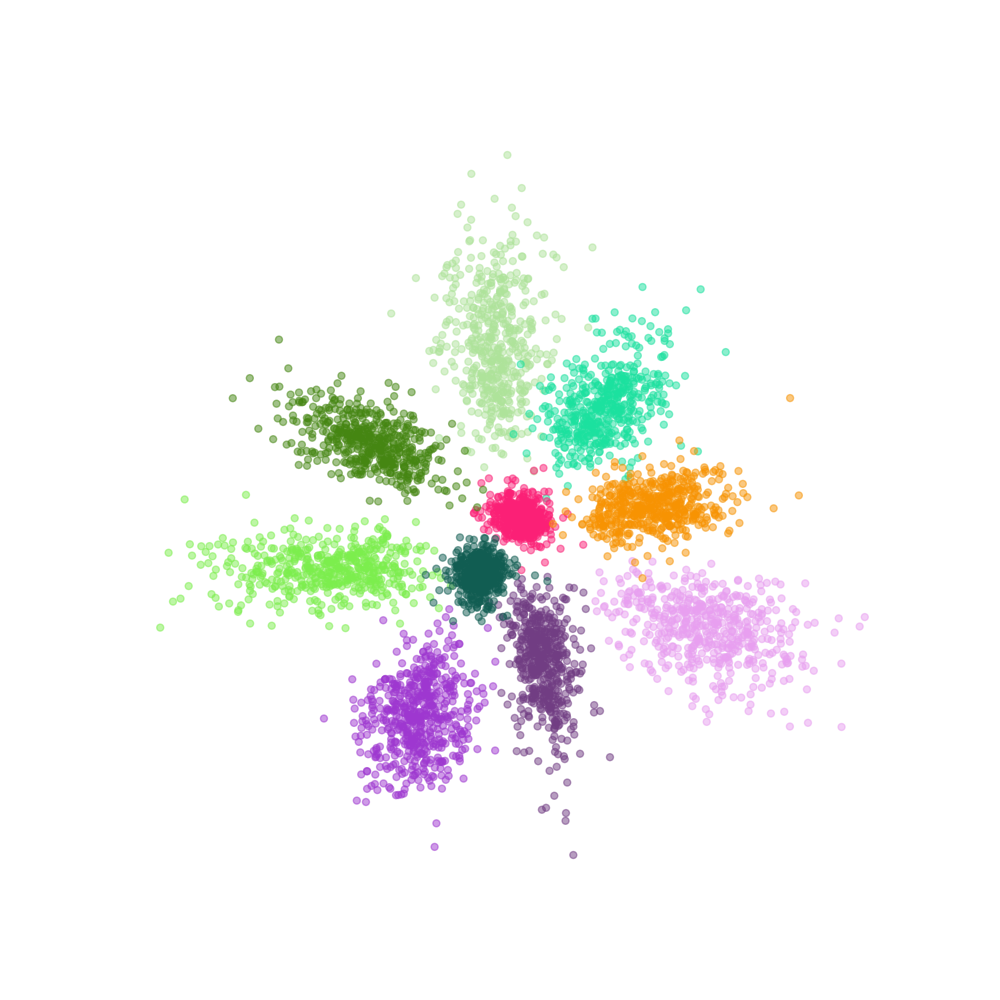}
        \caption{ShaRP (0.97)}
        \label{fig:inset-shrinp-proj}
    \end{subfigure}
    \caption{Comparison of projections of the MNIST dataset learned using (a) Auto-encoders, (b) SSNP\,\cite{DBLP:conf/ivapp/EspadotoHT21}, and (c) ShaRP. SSNP and ShaRP were trained using the \emph{ground truth} labels as class information --- encoded, here and next, by colors. Values in brackets are Distance Consistency scores (DSC \cite{SipsDSCMetric}), a quality metric that measures separability of clusters, with 1 being a perfect score.}
    \label{fig:comparison-ae-ssnp-shrinp}
\end{figure}

ShaRP builds atop of the recent DR method SSNP\,\cite{DBLP:conf/ivapp/EspadotoHT21}. SSNP extends a vanilla auto-encoder with loss $\mathcal{L}_{\text{AE}}$ 
with a classifier head (with an accompanying loss $\mathcal{L}_{\text{class}}$), yielding the total loss to be optimized as
\begin{equation}
    \mathcal{L}_{\text{SSNP}}(\*X, \hat{\*X}, \bar{Y}, \hat{Y}) = \mathcal{L}_{\text{AE}}(\*X, \hat{\*X}) + \rho\mathcal{L}_{\text{class}}(\bar{Y}, \hat{Y}).
\end{equation}
The projection $\*r_i \in \mathbb{R}^2$ of each input $\*x_i$ is generated by the bottleneck layer of the network. The classification loss $\mathcal{L}_{\text{class}}$, together with target labels or \emph{pseudolabels} generated by a clustering algorithm, enables SSNP to separate data clusters better than plain auto-encoders (see \autoref{fig:comparison-ae-ssnp-shrinp}). Yet, as the figure shows, SSNP collapses some clusters into elongated shapes, which we argue is (a) unnatural, as it suggests some anisotropy in the sample distribution; (b) space-inefficient, as much white space is not used to depict data; and (c) suboptimal for visualization as we cannot \emph{e.g.} easily select a cluster by rubberband tools or annotate it with a square-like icon.

ShaRP overcomes these shortcomings of SSNP by an explicit user-controlled shape regularization mechanism, described next (see also \autoref{subsec:controlling-shapes} for examples). ShaRP replaces SSNP's Auto-Encoder (AE) with a Variational AE (VAE)\,\cite{DBLP:journals/corr/KingmaW13}. 
The key AE-VAE difference is the latter's use of a \emph{sampling} process in the network's bottleneck layer. This, coupled with a necessary KL-Divergence regularization term 
\begin{equation}
    \label{eqn:kld-general}
    \mathcal{L}_{\text{reg}}(\theta) = D_{\text{KL}}(q_\theta || p) \dot= \mathbb{E}_{\*z\sim q_\theta}[\log(q_\theta(\*z)/p(\*z))].
\end{equation}
has as an immediate effect on the regularization of the learned latent space: Using $\mathcal{L}_{\text{reg}}$ pushes the learned probability distributions $q_\theta$ toward a standard form $p$ defined a priori (e.g., a standard Gaussian distribution) which prevents learning degenerate distributions. Also, crucially for our goals, this loss can be \emph{exploited} to model different shape regularization constraints (see next \autoref{subsec:controlling-shapes}). The complete loss function for ShaRP then reads as

\begin{align}
    \mathcal{L}_{\text{ShaRP}}(\*X, \hat{\*X}, \bar{Y}, \hat{Y}, \Theta) =& \mathcal{L}_{\text{recon}}(\*X, \hat{\*X}) + \rho\mathcal{L}_{\text{class}}(\bar{Y}, \hat{Y}) + \beta\mathcal{L}_{\text{reg}}(\Theta) \label{eqn:sharp-loss} \\
    =& \mathcal{L}_{\text{SSNP}}(\*X, \hat{\*X}, \bar{Y}, \hat{Y}) + \beta\mathcal{L}_{\text{reg}}(\Theta), \nonumber
\end{align}
where we make the connection to the SSNP loss explicit.

By using a suitable \emph{sampling} process, the clusters emerging in the projection will be shape-regularized. For instance, using a 2D Gaussian sampling distribution  yields \emph{elliptical} shapes (see \autoref{fig:inset-shrinp-proj}) because the equidensity contours of a 2D Gaussian are ellipses. This is dependent on $\mathcal{L}_{\text{reg}}$ preventing the degenerate learning of low (respectively, high) variances, which would give rise to point-like (resp., line-like) shapes in the projection.

\vspace{-0.15cm}
\subsection{Controlling cluster shapes}
\label{subsec:controlling-shapes}

We use as regularization targets the following shapes.

\begin{figure*}[htbp]
    \centering
    \includegraphics[width=\linewidth]{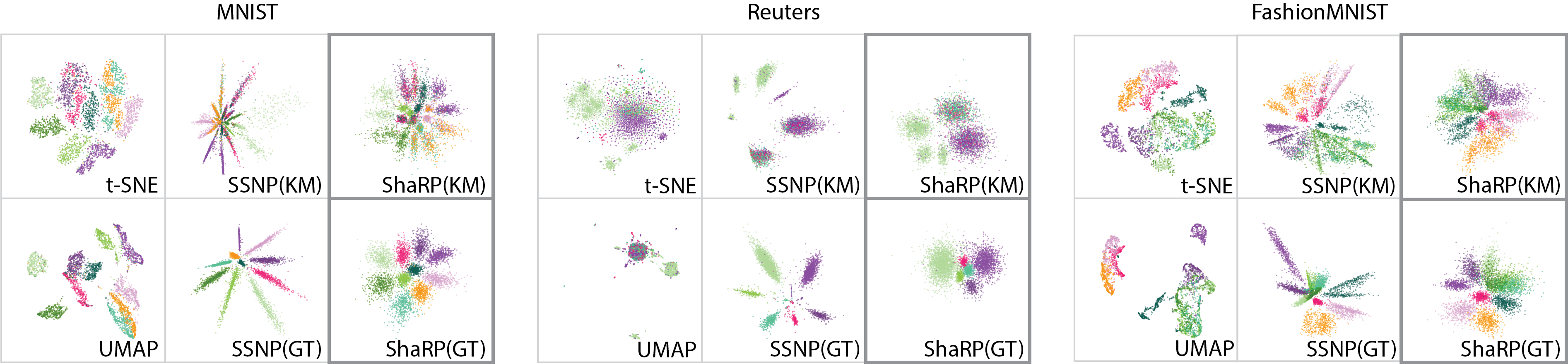}
    \caption{Our ShaRP method produces cluster shapes regularized towards a user-chosen target --- here, ellipses --- and can handle diverse data distributions. We demonstrate this here for the cases where we use ground truth labels (GT) or K-Means-generated pseudolabels (KM). We compare our results to SSNP (GT, KM) and to t-SNE and UMAP. More comparisons are present in the supplemental material.}
    \label{fig:RedHerrings}
\end{figure*}

\smallskip
\noindent\textbf{Ellipses.} Consider a diagonal Multivariate Normal distribution, \emph{i.e.}, $\*z_i \sim \mathcal{N}(\vec\mu, \operatorname{diag}(\vec\sigma^2))$. The natural prior to use here is the standard Multivariate Normal distribution $\mathcal{N}(\vec{\*0}, \*I)$ which simplifies sampling, propagating gradients, and calculating the KL-Divergence loss --- see \cite{DBLP:journals/corr/KingmaW13} for more details. By using this prior, we encourage learned probability distributions to be as close as possible to a standard Gaussian. Hence, the learned projection will output data clusters that resemble circles or ellipses (see \autoref{fig:RedHerrings}).

Using a Gaussian sampling distribution is standard for VAEs. For our projection goals, tweaking the sampling distribution and using suitable priors allows favoring different cluster shapes. We can use \emph{any} distribution, as long as we have \begin{enumerate*}[label=(\roman*)] \item access to log-probabilities of samples under the learned distribution and the prior; \item a way to propagate gradients through the sampling procedure (using a reparametrization trick or otherwise). \end{enumerate*} 

Access to the log-probabilities of samples under learned distributions and the prior removes the (constraining) need to analytically calculate the KL-Divergence since we can re-express it as a sample-based computation as
\begin{align}
    D_{\text{KL}}(q_\theta || p) %
    \approx \frac{1}{m} \sum_{i=1}^m \left(\log q_\theta(\*z_i) - \log p(\*z_i) \right),
    \label{eqn:kld-sampled}
\end{align}
where the approximation holds if $\*z_i \sim q_\theta(\cdot)$. Our next examples of regularization shape targets use Equation~\ref{eqn:kld-sampled} of computing the (approximate) KL-Divergence.

\smallskip
\noindent\textbf{Rectangles.} 
To create rectangular clusters, we use a generalized Normal ($\mathcal{GN}$) probability distribution. It introduces an additional shape parameter to the Gaussian (here denoted $\omega$) and has a density function of the form
$$
p(x | \mu, \alpha, \omega) \propto \exp\left(- (|x - \mu| / \alpha)^\omega\right).
$$

\begin{figure}[htbp]
    \centering
    \includegraphics[width=.7\linewidth]{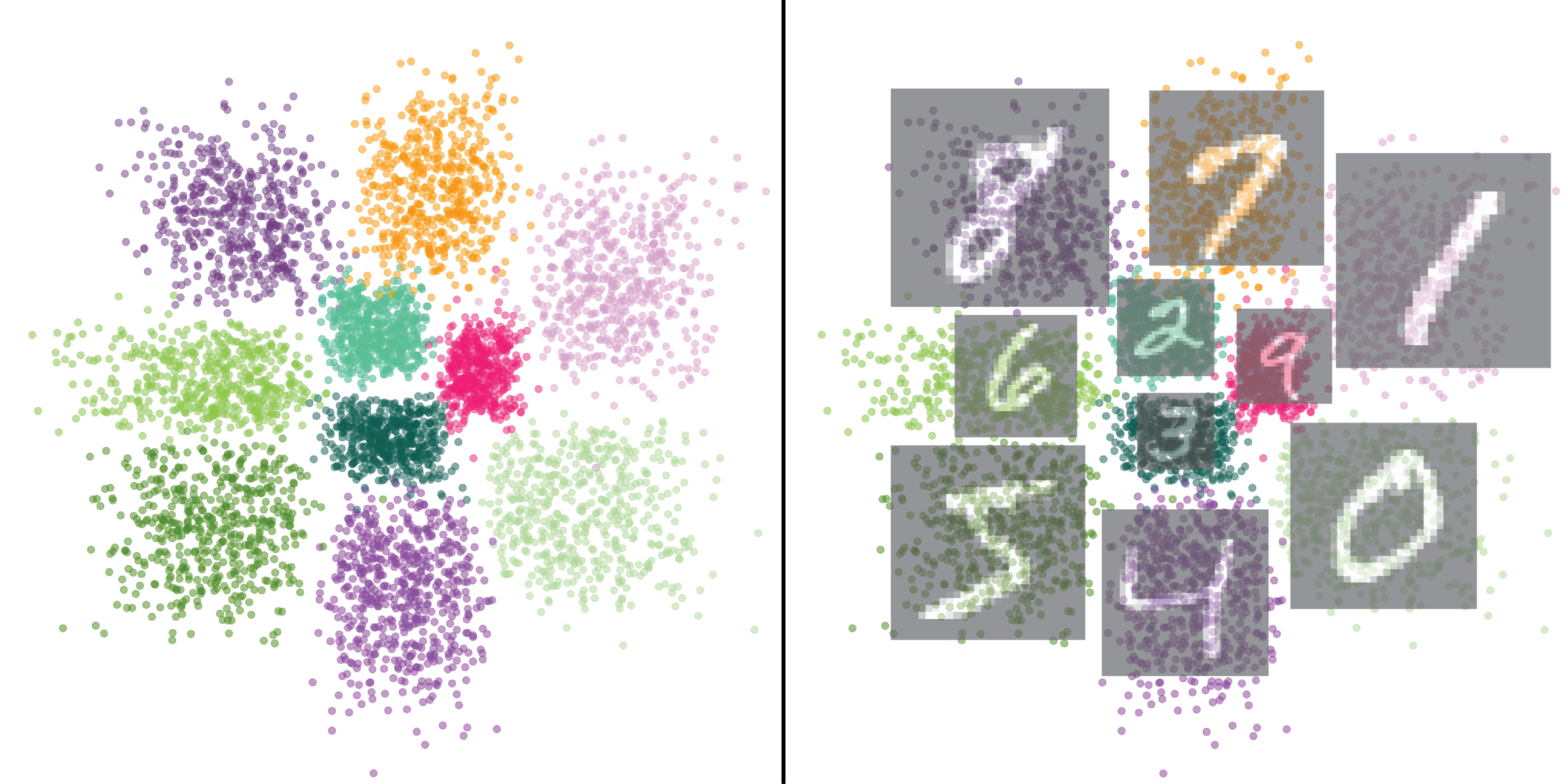}
    \vspace{-0.1cm}
    \caption{Shaping clusters as rectangles can be convenient for data labeling tasks, as illustrated by the right image where class image representatives are overlaid atop their respective clusters. We achieve this using a Generalized Normal distribution for sampling, here shown on the MNIST dataset for $\omega=10$ (left).}
    \vspace{-0.1cm}
    \label{fig:squaredness-overlaid-mnist}
\end{figure}

Tuning $\omega$ makes the tails of the distribution heavier or lighter. This is similar to the Minkowski p-norm where higher $p$ values (analogous to $\omega$) make sets of equidistant points approach axis-aligned squares as $p \to \infty$ instead of circles ($p = 2$). Using this distribution for sampling, with a high $\omega$ value, yields cluster shapes that resemble squares/rectangles instead of ellipses (see \autoref{fig:squaredness-overlaid-mnist}). %

\smallskip
\noindent\textbf{Convex polygons.}
If $\*V \in \mathbb{R}^{2 \times v}$ is a matrix of a base convex polygon's $v$ vertices in $\mathbb{R}^2$ and $\*w \in [0,1]^v$ is a vector such that $w_i \geq 0~\forall i,\, \sum_{i=1}^v w_i = 1$, then $\*p = \*V\*w$ is a point inside the base polygon with barycentric coordinates $w_i$. To sample points inside this polygon, we 
use the Dirichlet probability distribution 
\begin{align*}
    \*w \sim \operatorname{Dir}&(\alpha_1, \alpha_2, \dots, \alpha_v) \Rightarrow \*w \in [0, 1]^v, ~~~\sum_{i=1}^v w_i = 1 \quad
    (\alpha_i > 0,~ \forall i)
\end{align*}
which generates vectors with the same properties as $\*w$ above.

\begin{figure}[thbp]
    \centering
    \begin{subfigure}{.32\columnwidth}
        \includegraphics[width=\linewidth, trim=200 200 200 200, clip]{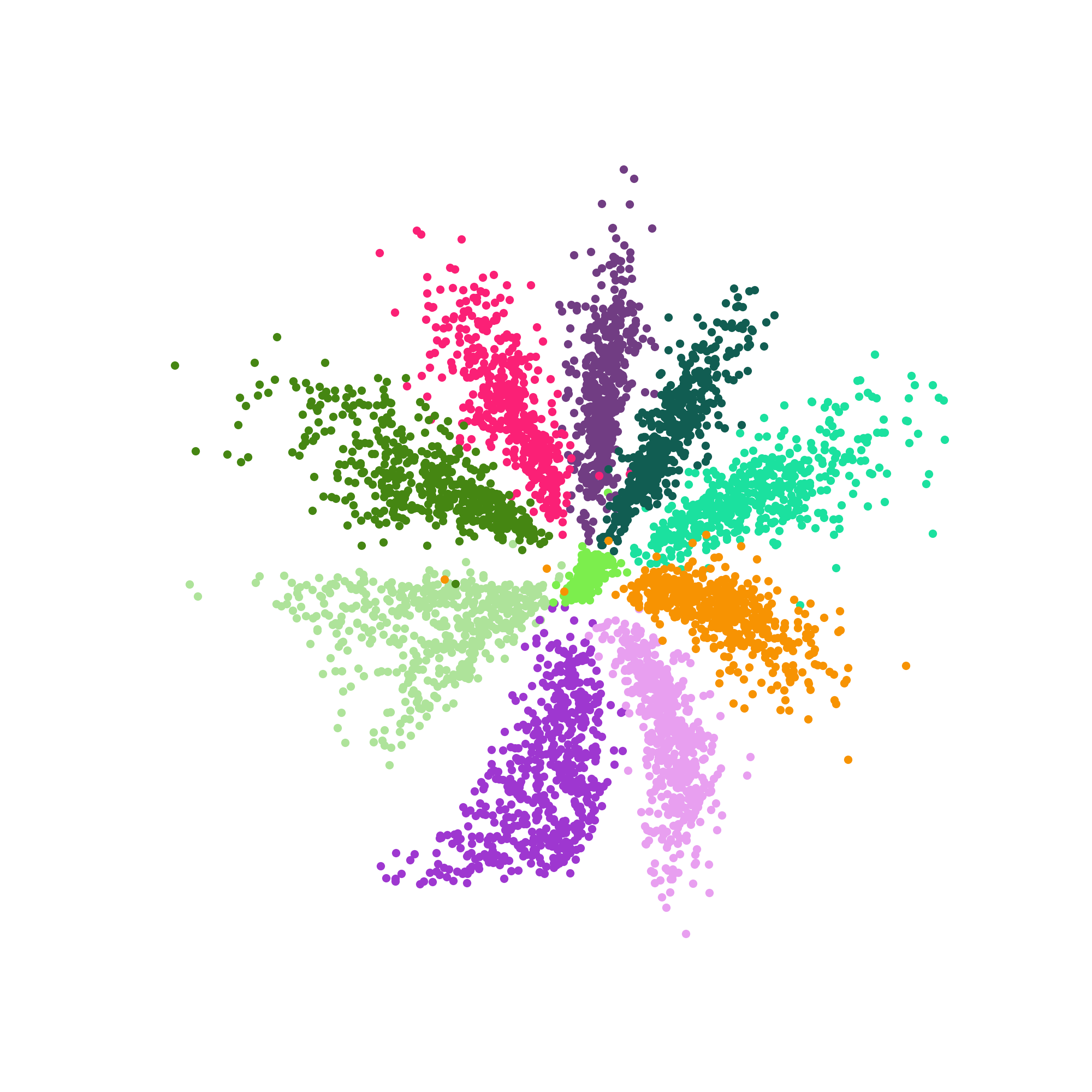}
        \caption{MNIST (0.905)}
        \label{fig:triangle-mnist}
    \end{subfigure}
    \begin{subfigure}{.32\columnwidth}
        \includegraphics[width=\linewidth, trim=200 200 200 200, clip]{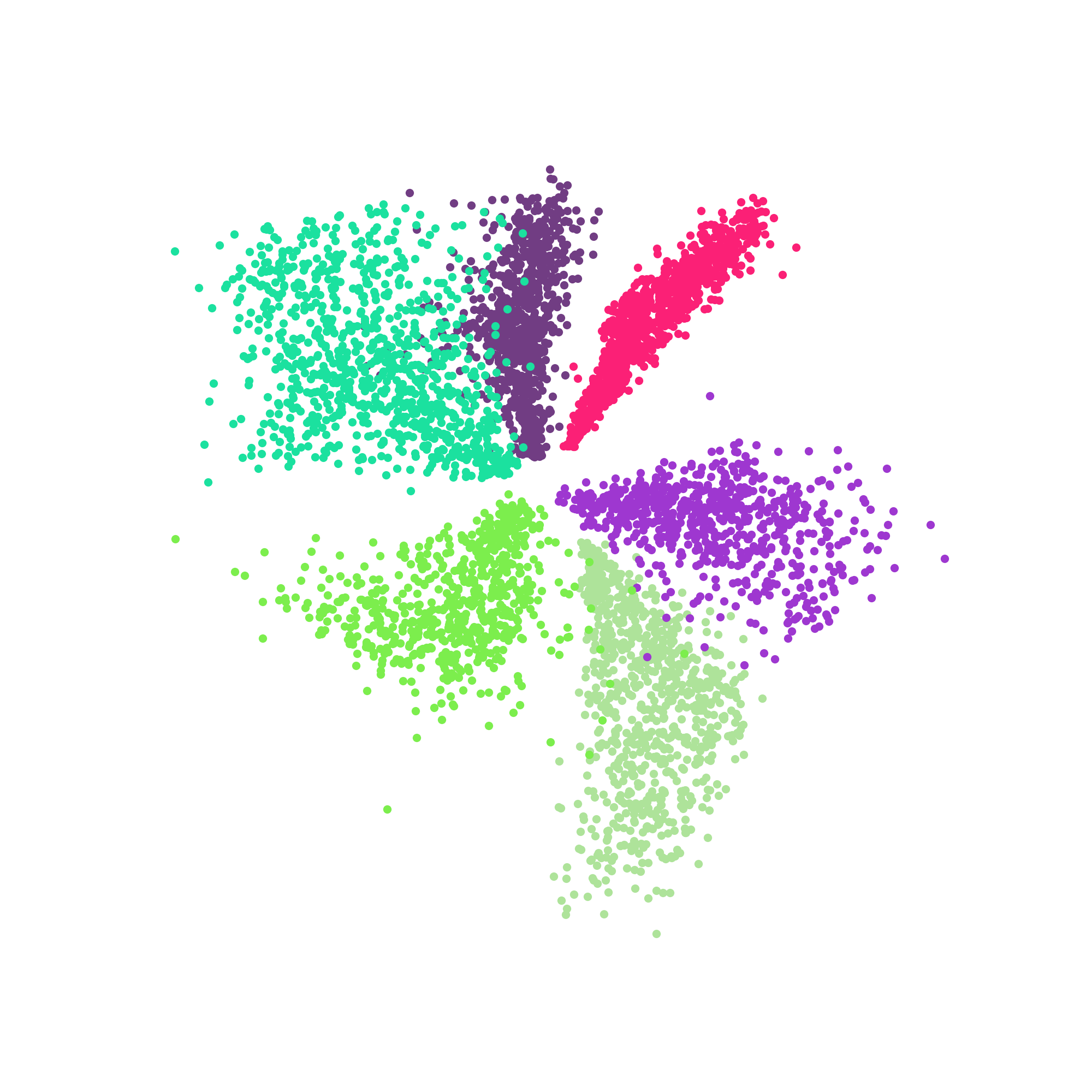}
        \caption{HAR (0.965)}
        \label{fig:triangle-har}
    \end{subfigure}
    \begin{subfigure}{.32\columnwidth}
        \includegraphics[width=\linewidth, trim=200 200 200 200, clip]{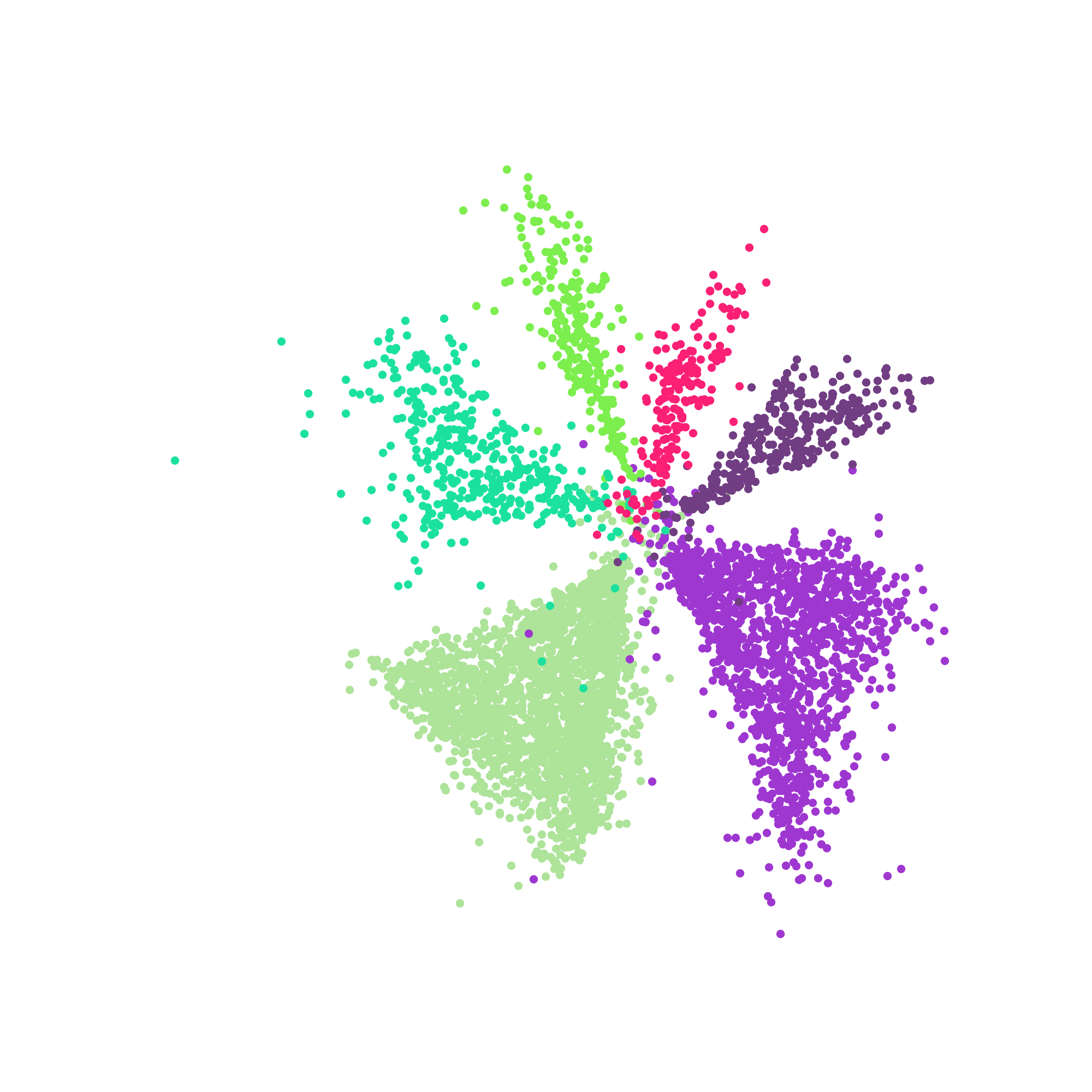}
        \caption{Reuters (0.959)}
        \label{fig:triangle-reuters}
    \end{subfigure}
    \caption{The results of our Triangular shaping sampling scheme over 3 different datasets. DSC values (in brackets) are close to the best value possible, indicating that we do not harm class separability.}
    \label{fig:triangle}
    \vspace{-0.1cm}
\end{figure}

This sampling scheme alone is not enough to learn a useful embedding since all data points will draw samples from the same region in space. Hence, we augment this scheme with rotation, scaling, and translation. \autoref{fig:triangle} shows this scheme for triangles, \emph{i.e.} $v = 3$, using as prior the ``uniform'' distribution on the triangle, which corresponds to $\operatorname{Dir}(1, 1, 1)$. 

\autoref{tab:distribution-per-shape} summarizes our proposed mechanisms for controlling cluster shapes by sampling distributions. For more technical details, see the supplemental material.

\begin{table}[thbp]
    \centering
    \small
    \caption{Correspondences between sampling schemes and shapes.}
    \begin{tabular}{ll|c}
       \textbf{Sampling} & \textbf{Prior} & \textbf{Shape} \\ \hline
       $\*z \sim \mathcal{N}(\mu, \operatorname{diag}(\sigma^2))$  & $\mathcal{N}(0, I)$ & $\bigcirc$ \\ \hline
       $\*z \sim \mathcal{GN}(\mu, \alpha, \omega)$  & $\mathcal{GN}(0, 1, \omega)$ & $\square$ \\  \hline
      
       $\*z \sim \operatorname{Dir}(\alpha_1, \alpha_2, \alpha_3)$ & \multirow{2}{*}{\vspace{-1.8em}$\operatorname{Dir}(1,1,1)$} & \multirow{2}{*}{\vspace{-1.8em}$\triangle$} \\
        $\*z \mapsto \begin{bmatrix}
            \cos\phi & -\sin\phi \\
            \sin\phi &  \cos\phi
        \end{bmatrix} \begin{bmatrix}
            s_x & 0 \\
            0 & s_y
        \end{bmatrix} \*V\*z + \begin{bmatrix}
            t_x \\ t_y
        \end{bmatrix}$
     &  &  \\
    \end{tabular}
    \label{tab:distribution-per-shape}
\end{table}

\vspace{-0.15cm}
\section{Evaluation}
\label{sec:Evaluation}

We next discuss how ShaRP gives direct control over cluster shapes while learning to project data (\autoref{subsec:eval-regularizing-shapes}), the quality of ShaRP projections (\autoref{subsec:eval-proj-quality}), and how tuning a single hyperparameter controls the shape regularization strength (\autoref{subsec:shape-regularization-intensity}). Finally, we discuss ShaRP's computational scalability  (\autoref{subsec:performance-metrics}).

\noindent\textbf{Datasets.} We use 5 datasets for evaluation (\autoref{tab:datasets}) which have different levels of classification difficulty, dimensionality, data type (images, motion data, text), and are often used in DR evaluations\,\cite{EspadotoDRSurvey}.

\begin{table}[htbp]
    \centering
    \scriptsize 
    \caption{Datasets used in our evaluation. \label{tab:datasets}}
    \begin{tabular}{lrr}
        \hline
        \textbf{Dataset} &  \textbf{Dimensionality ($n$)} & \textbf{\# classes ($K$)} \\ \hline
        \multicolumn{1}{l|}{USPS \cite{uspsdataset}} & 256 & 10 \\
        \multicolumn{1}{l|}{HAR \cite{AnguitaHARDataset}} & 561 & 6 \\
        \multicolumn{1}{l|}{MNIST \cite{MNISTLeCun}} & 784 & 10 \\
        \multicolumn{1}{l|}{FashionMNIST \cite{DBLP:journals/corr/abs-1708-07747}} & 784 & 10 \\
        \multicolumn{1}{l|}{Reuters \cite{Thom2017-reuters}} & 5000 & 6\\
        \hline
    \end{tabular}
\end{table}

\noindent\textbf{Techniques.} We compare ShaRP with t-SNE, UMAP, and Isomap, due to their wide adoption in the DR arena. We also compare with Auto-Encoders since they are a key building block of our technique; with SSNP since we are extending it; and with NNP\,\cite{EspadotoNNP}, a technique that learns to imitate projections, here trained to imitate t-SNE. For both ShaRP and SSNP, we use three different label sources: (1) from the ground truth of the dataset (GT); and pseudolabels created by the K-Means (2, KM)\,\cite{DBLP:journals/tit/Lloyd82} and Agglomerative (3, AG)\,\cite{DBLP:books/wi/KaufmanR90} clustering techniques.

\noindent\textbf{Hyperparameter settings.} We train ShaRP with the Adam optimizer using default parameter settings. We add L2 regularization to the bottleneck layer of the network with a coefficient of $0.5$. We use $\rho = 1$ and $\beta = 0.1$ and train using mini-batches of 256 data points.

\subsection{Generating shape-regularized projections}
\label{subsec:eval-regularizing-shapes}

\autoref{fig:RedHerrings} shows examples of how ShaRP can regularize learned projections. Instead of producing scatterplots where cluster shapes, sizes, and intercluster spacing are widely different (as with t-SNE and UMAP), ShaRP generates a more similar representation of the high-dimensional data in each 2D projection
(intra-projection regularization). Also, the visual signature obtained is consistent throughout datasets (inter-projection regularization). The learned projections do well with respect to quality metrics (see \autoref{tab:proj-quality-metrics} and its discussion in \autoref{subsec:eval-proj-quality}). All images were generated using the same hyperparameter values, which shows the robustness of ShaRP to different datasets.

\subsection{Measuring the projection quality}
\label{subsec:eval-proj-quality}

We evaluate ShaRP by a set of established projection quality metrics (trustworthiness, continuity, Shepard correlation, normalized stress, neighborhood hit, and distance consistency) following Espadoto et al.\,\cite{EspadotoDRSurvey,DBLP:conf/ivapp/EspadotoHT21}.
Precise metric definitions are listed in the supplemental material. 
We compute metrics over all datasets using a Gaussian sampling layer which produces ellipse-like clusters.
\autoref{tab:proj-quality-metrics} shows these mean and standard deviations of the metrics over all datasets for ShaRP and the other six evaluated techniques. 
We see that ShaRP avoids very high Stress values (present in t-SNE, \emph{all} studied datasets; UMAP and AEs, some datasets). We do, however, have higher Stress than SSNP, since we \emph{force} clusters into desired shapes, which can require projected (2D) distances to be quite different from data-space distances. Given that our Stress is still lower than t-SNE, UMAP, and AE, we believe this is a reasonable trade-off. For the other metrics, ShaRP performs comparably to t-SNE and UMAP. Overall, we claim that ShaRP offers its capability of shape regularization without negatively impacting quality. It is worth noting that, for their AG and KM versions, both SSNP and ShaRP can be held back by the clustering algorithm's ability to properly group the dataset into classes.

To test how ShaRP's support of different regularization shapes affects projection quality, we asked ShaRP to produce clusters in five shapes -- ellipses (using Gaussian sampling); rectangles ($\omega = 5$ and $\omega = 15$, see \autoref{subsec:controlling-shapes}); and triangles (translated in projected space and respectively forced to $t_x = t_y = 0$), for all 5 tested datasets.
\autoref{tab:shapes-metrics} shows the mean and standard deviation of quality metrics per dataset. We see little variation in these metrics. This points to the robustness of ShaRP and further supports our claim that controlling the visual signatures of projections can be done without (strongly) influencing quality metric values.

\begin{table*}[htb]
    \centering
    \scriptsize
    \caption{Means and (standard deviations) of metrics computed for ShaRP and six other projection methods over 5 datasets. See \autoref{subsec:eval-proj-quality}. For non-aggregated results, see supplemental material. \label{tab:proj-quality-metrics}}
    \begin{tabular}{lrrrrrr}
    \hline
         Method & Trustworthiness & Continuity & Shepard Corr. & Stress & Neigh. Hit & Dist. Consistency \\ \hline
         \multicolumn{1}{l|}{Isomap} & 0.817 (0.126) & 0.927 (0.096) & 0.540 (0.410) & 4.355 (4.359) & 0.737 (0.091) & 0.587 (0.095) \\
\multicolumn{1}{l|}{t-SNE} & 0.940 (0.110) & 0.968 (0.032) & 0.436 (0.250) & 25.951 (18.373) & 0.914 (0.064) & 0.754 (0.141) \\
\multicolumn{1}{l|}{UMAP} & 0.910 (0.139) & 0.960 (0.055) & 0.424 (0.336) & 1.389 (1.451) & 0.874 (0.088) & 0.741 (0.166) \\
\multicolumn{1}{l|}{NNP[t-SNE]} & 0.903 (0.122) & 0.967 (0.033) & 0.441 (0.254) & 0.872 (0.044) & 0.859 (0.075) & 0.742 (0.141) \\
\multicolumn{1}{l|}{AE} & 0.878 (0.136) & 0.917 (0.091) & 0.342 (0.331) & 1.236 (1.075) & 0.793 (0.043) & 0.622 (0.082) \\
\multicolumn{1}{l|}{SSNP (AG)} & 0.849 (0.139) & 0.922 (0.080) & 0.461 (0.180) & 0.301 (0.067) & 0.812 (0.058) & 0.674 (0.090) \\
\multicolumn{1}{l|}{SSNP (KM)} & 0.862 (0.142) & 0.928 (0.065) & 0.451 (0.221) & 0.334 (0.134) & 0.777 (0.046) & 0.620 (0.104) \\
\multicolumn{1}{l|}{SSNP (GT)} & 0.797 (0.126) & 0.902 (0.078) & 0.454 (0.117) & 0.500 (0.061) & 0.977 (0.025) & 0.930 (0.043) \\ \hline
\multicolumn{1}{l|}{ShaRP (AG)} & 0.816 (0.136) & 0.886 (0.108) & 0.382 (0.311) & 0.770 (0.089) & 0.771 (0.047) & 0.661 (0.073) \\
\multicolumn{1}{l|}{ShaRP (KM)} & 0.832 (0.143) & 0.897 (0.082) & 0.426 (0.268) & 0.782 (0.079) & 0.747 (0.045) & 0.658 (0.051) \\
\multicolumn{1}{l|}{ShaRP (GT)} & 0.755 (0.118) & 0.864 (0.098) & 0.343 (0.221) & 0.783 (0.085) & 0.939 (0.056) & 0.890 (0.074) \\ \hline

    \end{tabular}
    
\end{table*}

\begin{table*}[htb]
    \centering
    \scriptsize
    \caption{Means and (standard deviations) of metrics for ShaRP computed for 5 datasets and 5 different settings for shape regularization. See \autoref{subsec:eval-proj-quality}. For non-aggregated results, see supplemental material. \label{tab:shapes-metrics}}
    \begin{tabular}{lrrrrrr}
        \hline
        Dataset  & Trustworthiness & Continuity & Shepard Correlation & Stress & Neigh. Hit & Dist. Consistency \\ \hline
        \multicolumn{1}{l|}{MNIST} & 0.735 (0.007) & 0.878 (0.035) & 0.167 (0.071) & 0.809 (0.073) & 0.965 (0.024) & 0.909 (0.068) \\
        \multicolumn{1}{l|}{FashionMNIST}  & 0.823 (0.014) & 0.888 (0.035) & 0.468 (0.079) & 0.837 (0.065) & 0.830 (0.018) & 0.761 (0.043) \\
        \multicolumn{1}{l|}{HAR} & 0.826 (0.001) & 0.841 (0.039) & 0.546 (0.053) & 0.690 (0.087) & 0.959 (0.021) & 0.925 (0.025) \\
        \multicolumn{1}{l|}{Reuters} & 0.555 (0.002) & 0.691 (0.011) & 0.302 (0.050) & 0.631 (0.083) & 0.969 (0.007) & 0.903 (0.035) \\
        \multicolumn{1}{l|}{USPS}  & 0.805 (0.006) & 0.899 (0.036) & 0.275 (0.105) & 0.700 (0.104) & 0.965 (0.012) & 0.904 (0.053) \\
        \hline
    \end{tabular}
\end{table*}

\subsection{Control of shape regularization intensity}
\label{subsec:shape-regularization-intensity}
\begin{figure}[thbp]
    \centering
    \includegraphics[width=\linewidth, clip, trim=15 0 15 0]{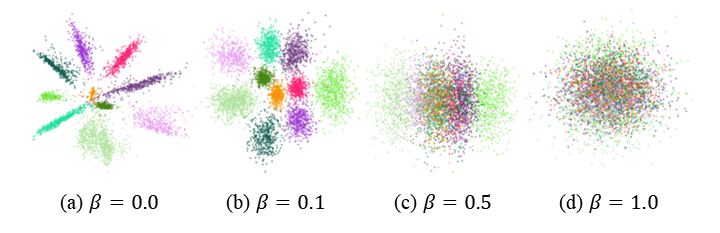}
\caption{The $\beta$ coefficient (\autoref{eqn:sharp-loss}) controls the shape regularization strength, shown here on the USPS dataset. $\beta = 0$ approximately reproduces SSNP (a). Increasing it (b, c) progressively forces the learned clusters into circular shapes, up to the point where they are no longer separable and the projection is of low quality (d).}
\label{fig:bubblyness-control}
\smallskip
    \begin{subfigure}{\linewidth}
        \centering
        \scriptsize
        \begin{tabular}{rrrrrrr}
            \hline
             $\beta=$ & $0.0$ & $0.05$ & $0.1$ & $0.25$ & $0.5$ & $1.0$ \\ \hline
             \multicolumn{1}{l|}{Trustworthiness} & 0.87 & 0.81 & 0.79 & 0.77 & 0.66 & 0.54 \\
             \multicolumn{1}{l|}{Continuity} & 0.95 & 0.93 & 0.85 & 0.74 & 0.70 & 0.56 \\ 
             \multicolumn{1}{l|}{Shepard Corr.} & 0.39 & 0.30 & 0.27 & 0.26 & 0.21 & 0.09 \\
             \multicolumn{1}{l|}{Stress} & 0.93 & 0.84 & 0.78 & 0.66 & 0.57 & 0.54 \\
             \multicolumn{1}{l|}{Neighborhood Hit} & 0.99 & 0.99 & 0.97 & 0.91 & 0.60 & 0.45 \\ \hline
        \end{tabular}
        \caption{Quality Metrics for varying values of $\beta$.}
    \end{subfigure}

\end{figure}

We adjust the amount of shape regularization through the $\beta$ multiplier in the loss function (\autoref{eqn:sharp-loss}). \autoref{fig:bubblyness-control} shows this: Larger $\beta$ values force clusters to conform to the shape generated in the sampling layer --- ellipses, in this case. Exaggerated shape regularization (high $\beta$), however, makes ShaRP favor `shape over data' too much and creates projections which cannot properly depict data --- sampling from a distribution similar to the prior overshadows producing a sensible embedding. In our tests, we have found a value of $\beta = 0.1$ to give consistently good results.

\subsection{Computational performance}
\label{subsec:performance-metrics}

\autoref{fig:performance-metrics} shows how ShaRP fares compared to other projection techniques \emph{vs.} computational time. Tests were run on a PC with an AMD Ryzen 9 5900HX 3.3GHz 8-core processor and an NVIDIA RTX 3080 GPU. ShaRP is much faster than t-SNE (50-80\% speedup) and Isomap (50-60\%). It is also faster than AE, UMAP, and only slightly slower than SSNP, its predecessor. The used batch size (256 data points) is largely responsible for ShaRP's speed. Also, since ShaRP has out-of-sample ability, we can train it on a representative data subsample to next project an entire dataset with high quality.

\setcounter{figure}{5}
\begin{figure}[htbp]
    \centering
    \includegraphics[width=.95\linewidth]{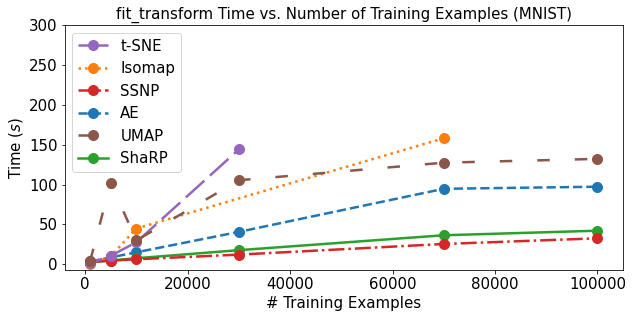}
    \caption{Run times for five projection techniques and ShaRP, (upsampled) MNIST dataset. Runs were stopped after 5 minutes. We see improvements ranging from 50-80\% with respect to t-SNE, UMAP, and Isomap, and similar performance to AE and SSNP, as expected. More results are available in the supplemental material. 
    \vspace{-0.15cm}
    \label{fig:performance-metrics}}
\end{figure}

\section{Discussion and Future Work}
\label{sec:Discussion}

ShaRP introduces a novel level of pattern steerability for a projection algorithm, all while performing comparably to state-of-the-art methods in relevant (visual) quality metrics. However, ShaRP also has some limitations which frame our future work directions.
Currently, we only support numerical features as these work directly with Auto-Encoders. One-hot encoding or Categorical Variational Auto-Encoders\,\cite{morton2021scalable} can overcome this limitation with only slight adaptions to our network architecture.
Also, the sampling schemes we devised support shaping clusters into ellipses and convex polygons --- with two different possibilities for shaping clusters into rectangles. A wider variety of shapes can be obtained by devising new sampling schemes. However, obtaining log-probabilities for samples of complex sampling schemes can be computationally intensive. We next aim to study more sampling schemes that naturally encourage further visual aspects of interest of projections, \emph{e.g.}, cluster separability. 

Shape-regularized projections also can help  interaction and visualization tasks. For example, rectangular shaped clusters can help as a clutter-reduction mechanism whenever (annotation) overlays with text and images should be added to projection (cf Fig.~\ref{fig:squaredness-overlaid-mnist}). We aim to further study this aspect, including the ease of interactive hierarchical navigation of thumbnail-annotated squarified projections to support various analysis tasks of high-dimensional data.

\newcommand{\etalchar}[1]{$^{#1}$}

\end{document}